\documentstyle[psfig,amsmath,eqsecnum,multicol,aps,prb]{revtex}

\title{Effect of boundary conditions on diffusion in two-dimensional 
granular gases}
\author{C. Henrique$^1$, G. Batrouni$^2$ and D. Bideau$^1$}
\address{$^1$ Groupe Mati\`ere Condens\'ee et Mat\'eriaux 
Universit\'e de Rennes I, B\^at 11A, 263, avenue du g\'en\'eral
Leclerc CS 74205. 35042 Rennes Cedex, France}
\address{$^2$ Institut Non-Lin\'eaire de Nice, Universit\'e de 
Nice-Sophia Antipolis, 1361 route des Lucioles, 06560 Valbonne,
France}

\begin{document}

\maketitle
\begin{abstract}
We analyze the influence of boundary conditions on numerical
simulations of the diffusive properties of a two dimensional granular
gas.  We show in particular that periodic boundary conditions
introduce unphysical correlations in time which cause the coefficient
of diffusion to be strongly dependent on the system size.  On the
other hand, in large enough systems with hard walls at the boundaries,
diffusion is found to be independent of the system size. We compare
the results obtained in this case with Langevin theory for an elastic
gas. Good agreement is found. We then calculate the relaxation time
and the influence of the mass for a particle of radius $R_s$ in a sea
of particles of radius $R_b$. As granular gases are dissipative, we
also study the influence of an external random force on the diffusion
process in a forced dissipative system. In particular, we analyze
differences in the mean square velocity and displacement between the
elastic and inelastic cases.

\end{abstract}

\section{Introduction}

The interactions among grains and between grains and the boundaries
influence profoundly the macroscopic behavior of granular systems.  To
study such complex many body systems, numerical simulations are
frequently used where one of the most important ingredients are the
collision laws introduced to treat interactions~\cite{schafer}.  For
dilute assemblies of grains one can use molecular dynamics algorithms
where periodic boundary conditions are usually used. If the system is
initially in a square box, a particle going out on the left re-enters
the system on the right.  We will show below that this kind of
boundary condition modifies the general dynamics of the grains and
introduces large correlations in time. This changes the diffusive
behavior of the grains.  In this paper, we propose an alternative
approach to calculate numerically the coefficient of diffusion,
accurately and with only very small finite size effects. To validate
our methods, we compare our results for an elastic gas with the
Langevin theory.

As granular gases are dissipative it is necessary to feed energy into
the system to keep the particles agitated. To thermalize the system,
we choose a random acceleration added to each grain at regular time
step intervals $dt$.  Our final goal in this paper is to study the
dependence of the dynamic properties of the granular gas on the mode
used to force the system. This work is a first step towards
understanding the diffusion process in a binary system composed of two
grain sizes.  The system considered here is composed of one particle,
{\bf s}, of radius $R_s$ in a sea of particles of radius $R_b$.  The
particles are spheres constrained to move in a plane and which
interact along their equators so that the system is two
dimensional. The system considered here is dilute with a packing
fraction of 30 \%.  The simulations are done with the molecular
dynamics algorithms ({\it time step driven}~\cite{allen} and {\it
event driven}~\cite{haile}).

To characterize the diffusive behavior, we focus on the mean square
displacement of the {\bf s} particle.  It is well known that for a 2D
gas, the integral of the auto-correlation function does not
converge~\cite{ernst}. This means that the mean square displacement
does not vary linearly with time.  Therefore, strictly speaking, we
cannot define a diffusion coefficient in 2D. However, we show that in
a limited range of time, in the stationary state, the mean square
displacement can be approximated by the linear function:
\begin{equation}
\label{eq:D}
<(\vec{r}(t+t_0)-\vec{r}(t_0))^2> \propto 4Dt
\end{equation}
where $D$ can be interpreted as a diffusion coefficient.  All
quantities are expressed in arbitrary units.

\section{Choice of boundary conditions}

In this section we show that periodic boundary conditions introduce
strong correlations and therefore alter the diffusion process.

\subsection{Periodic boundary conditions} 

Consistent with common practice, we have used periodic boundary
conditions to simulate a system of identical spheres $R_s=R_b=0.5$.
Initially the particles are placed randomly in a square box of length
$L$. The number of particles is calculated for each system depending
on $L$, $R_s$ and the packing fraction.  Periodic boundary conditions
are applied in both directions.  In this case, for elastic or forced
gases (section~\ref{therm}), we have observed a strong dependence of
$D$ (or of the mean square displacement) on the system size.

In Fig.~\ref{fig:perio}, we have plotted the mean square displacement
$<(\vec{r}(t+t_0)-\vec{r}(t_0))^2>$ (Fig.~\ref{fig:perio}a) and $\int
C(t) dt$ (Fig.~\ref{fig:perio}b), both calculated in the stationary
state, as function of $t$. $C(t)$ is the normalized autocorrelation
function:
\begin{equation}
\label{eq:c-t}
C(t)=\frac{<\vec{\mathrm{v}}(t_0+t)\vec{\mathrm{v}}(t_0)>
-<\vec{\mathrm{v}}(t_0)>^2} {<\vec{\mathrm{v}}(t_0)^2>
-<\vec{\mathrm{v}}(t_0)>^2}.
\end{equation}
First, we note that the mean square displacement, at large time,
varies linearly with time as expected but the slope of the curve,
i.e. the diffusion coefficient, increases with system size. We show,
in the inset to Fig.~\ref{fig:perio}a, that this dependence on $L$
appears already at short time, when $<(r^2(t+t_0)-\vec{r}(t_0))^2> \ll
L^2$. This feature can be also observed in {\small $\int\limits_{0}^{t
\sim \infty}C(t)dt $}, which is proportional to the diffusion
coefficient. Similarly, we observe that the relaxation time $\tau_r$
(i.e. $C(\tau_r) \simeq 0 $) increases with size.  In summary, the
bigger the system is, the longer the characteristic time $\tau_r$ and the
larger the diffusion coefficient $D$ are.  We recall that such dependence
has been observed by Alder {\it et al}~\cite{alder1}. They proposed
the following law for the dependence of $D$ on the number of
particles, $N$,
\begin{equation}
D(N) = D(\infty) (1 -2/N).
\end{equation}
However their numerical simulations do not support this
conjecture~\cite{alder2} since they fail to observe any saturation of $D$ for large
systems. In addition, they found strong correlations in the velocity
field characterized by the presence of vortex flow pattern at the
microscopic scale. Our results confirm the lack of convergence for
$D$ with system size. In addition, this variation of $D$ with $L$ is
also observed in the case of inelastic collisions. 

Another important remark is in order.  If the system size is, for
example, 60 (with 1400 particles of radius $R=0.5$), the
characteristic time $\tau_r$ is found to be around $20$ which
represents about 200 collisions for a particle. This means that a
particle needs to undergo 200 collisions to lose completely the memory
of its past. According to the Boltzmann theory this time should be
limited to only a few collisions.  Therefore we cannot accept this
result as a valid macroscopic description of a gas. It is worth noting
that the same results are found for both, the time step driven and
the event driven algorithms.

We now discuss some points helpful for understanding the problem.
Initially, each particle has a random velocity drawn from a Maxwellian
distribution. We shift the linear and angular momenta so that the
system has zero center of mass momentum and zero angular momentum
relative to the center of mass.  We find, however, that, although the
system keeps its center of mass at rest throughout the simulation, the
system is no longer isotropic, its moment of inertia becoming that of
an ellipsoid. Let $I(t)$ be the inertia matrix of the system. Its two
eigenvalues $\lambda_n$ and $\lambda_p$ are related via
\begin{equation}
\lambda_n + \lambda_p = m\sum\limits_{k=1}^{N}r_k^2(t),
\end{equation}
where the sum is over all $N$ particles each of mass $m$. Following
$\lambda_n$ and $\lambda_p$ in time shows that the system takes an
ellipsoidal form ($\lambda_n < \lambda_p$).  We have found, as well,
an anisotropy in the diffusion tensor $\Hat{D}(t)$ defined from $I(t)$
as:
\begin{equation}
\Hat{D}(t)=\frac{1}{Nm}\frac{I(t+ \delta t) - I(t)}{\delta t}
\end{equation}
As example we show, in Fig.~\ref{fig:D1D2} the two eigenvalues $D_1$
and $D_2$ of $\Hat{D}$ as functions of $t$ for a particular periodic
system. Clearly, $D_1$ and $D_2$ are very different for all $t$. For
all systems we studied, we found two different diffusion coefficients
which depend strongly the system size. We were not able to find how
these values scale with $L$.

In addition, we have found that, contrary to its initial condition,
the system starts to rotate. This fact is put in evidence by
calculating the two eigenvectors $\vec{u}_n$ and $\vec{u}_p$ of
$I(t)$. These two (perpendicular) vectors rotate in space and, most
importantly, they keep the same direction of rotation for a long time
($\sim \tau_r$). We suspect that this rotation induces an anomalous
temporal correlation of velocities. One should point out that this
rotation phenomenon seems similar to that observed by Alder {\it et
al} in their simulations with similar periodic boundary conditions.

We strongly believe that the use of periodic boundary conditions is
responsible for this anomalous correlation.  These boundary conditions
present another inconvenience which is connected with the rotation of
the system: The square geometry of the system does not permit the
conservation of distances between two particles when the system is
rotating. In Fig.~\ref{fig:iandj} we show that after a rotation of
$\theta$ the distance $d_{ij}$ between particles $i$ and $j$ can be
drastically changed by the rotation if one of them goes through the
boundary.  Because the distance between particles is not conserved by
rotation, the interaction potential used in the algorithm, which
depends only on the relative positions of particles $d_{ij}$, is
itself not invariant under rotation and so the angular momentum of the
system is not conserved. Effectively the total angular momentum is
fluctuating as one can see in Fig.~\ref{fig:lz-t}.  Every time a
particle goes through the boundary, its angular momentum, $l^i_z$,
changes sign. Consequently, the change in angular momentum is $\Delta
L_z= -2l^i_z$. $\Delta L_z$ is always proportional to $L$ (the system
size) and the total number of particles, $N$, is proportional to
$L^2$. However it appears that the fluctuations of $L_z$ get bigger
with system size (see Fig.~\ref{fig:lz-t}).

The use of periodic boundary conditions amounts to replicating the
system on a square lattice. There are, therefore, several identical
systems which interact through the boundaries. The rotation observed
in our system is then extended to all these systems and can create
some shear stress, due to frustration of rotation, between neighboring
systems.

These boundary conditions can have other consequences on the dynamics
of granular systems. For example, during the simulation of a cooling
state the system evolves towards clusters~\cite{gold} whose
orientation depends on the type of boundary conditions~\cite{gold2}.
In other similar simulations~\cite{bizon} it was shown that there
exist large spatial correlations between particles where the
velocities stay correlated over a distance of about $L/2$.

\subsection{Reflecting boundaries}

\label{sec:reflected}

In the case of reflecting boundaries the system is rotationally
invariant leading to better behavior of the mean square
displacement. However, in this case, the mean square displacement is
limited at long time by the system size. To circumvent this problem,
we proceed as follows. The test particle {\bf s} is initially put at
the center of system at $t=0$. The evolution of the position and
velocity of this test particle are then followed until it reaches the
boundary of the system in time $t_w$. We then repeat this many times
collecting statistics for many test particles with different initial
velocities.

The mean square displacement is calculated over 500 such trajectories
and limited to time smaller than the smallest $t_w$.  In this case, as
one can see in Fig.~\ref{fig:R2_wall}, there is no depen\-dence of the
mean square displacement on the system size. Therefore, we can now
trust the results of our numerical simulations. 

We recall that the integral of the velocity correlation function does
not converge in 2D. However, in a limited range of time (see
Fig.~\ref{fig:R2_wall}), the quantity
$<(\vec{r}(t+t_0)-\vec{r}(t_0))^2>$ can be approximated by a straight
line and $D$ calculated according to Eq.~(\ref{eq:D}). Therefore, the
estimate of $D$ with this method is an approximation.

\section{Diffusion in an elastic gas}

We first validate our algorithm using reflecting boundaries for an
elastic gas (i.e., where the collision between particles are elastic).
Then, we compare the numerical results with those given by the
Langevin equation. Indeed, near equilibrium, the dynamics of {\bf s}
can be described approximately, by a Langevin equation:
\begin{equation}
\label{eq:lang1}
\begin{array}{c}
\displaystyle{\frac{d \mathrm{v_i}(t)}{dt} = - \gamma \mathrm{v_i}(t) + \Gamma_i(t),} \\
<\Gamma_{i}(t)\Gamma_{j}(t')> = q\delta_{i,j}\delta(t-t').
\end{array}
\end{equation}
where $i$ denotes the two direction $x$ and $y$.
Integrating Eq.~(\ref{eq:lang1}), the dependence of the mean square velocity
on time is simply given by:
\begin{equation}
\label{eq:lang3}
v^2(t)=v^2(0) e^{-2\gamma t} + \frac{q}{\gamma}(1 - e^{-2\gamma t}).
\end{equation}
In this paper $\mathrm{v}$ denotes the instantaneous velocity of one
particle and $v^2$ the mean square velocity averaged over the
different {\bf s} trajectories.  We can rewrite Eq.~(\ref{eq:lang3})
using the mean square velocity in the equilibrium state $v^2(\infty)=
q/\gamma$:
\begin{equation}
\label{eq:lang4}
v^2(t)= v^2(\infty) + (v^2(0) - v^2(\infty))e^{-2\gamma t}.
\end{equation}
where $1/\gamma$ corresponds to the relaxation time.

For example, if $R_s>>R_b$ or, equivalently, $m_s>>m_b$ ($m_{s,b}$ is
the mass of the particle of radius $R_{s,b}$), $1/\gamma$ is very
large: the collision of {\bf s} with a light particle $b$ will not
affect strongly the velocity of {\bf s}. So a great number of
collisions is needed before {\bf s} reaches its equilibrium state.
Knowing the total kinetic energy of the system $E_k^{tot}$ (which is
given by the initial velocity of each particle), we can easily
calculate the square velocity in the equilibrium state $v^2(\infty)=
q/\gamma$ (using the Boltzmann distribution law for elastic
gases). Using the simulations to calculate $v^2(t)$ for the {\bf s}
particles for initial conditions which are very different from the
stationary state, {\it i.e.} $v^2(0)\ne v^2 (\infty)$ and comparing
with Eq.~(\ref{eq:lang4}), we obtain the relaxation time for a big
(heavy) particle (Fig.~\ref{fig:lange}).  Note that we are looking for
agreement near the equilibrium state, where Eq.~(\ref{eq:lang1}) is
valid. Indeed the dissipation term $\gamma$ must depend on both
velocities $v^2_s$ and $v^2_b$, as we will show.

Equation~(\ref{eq:lang1}) also gives the mean square displacement as a
function of time,
\begin{equation}
\label{eq:lang5}
<(\vec{r}(t) - \vec{r}(0))^2> = (v_0^2 - \frac{q}{\gamma})\frac{(1
-e^{-\gamma t})^2}{\gamma^2} + \frac{2q}{\gamma^2}t -
\frac{2q}{\gamma^3}(1-e^{-\gamma t}).
\end{equation}
Comparing Eq.~(\ref{eq:lang5}) and Eq.~(\ref{eq:D}) at large time, the
coefficient of diffusion is seen to be
$D=\frac{v^2(\infty)}{2\gamma}$.  In Fig.~\ref{fig:lange}, we compare
the theoretical mean square displacement, Eq.~(\ref{eq:lang5}), with
the numerical one, obtained for the case $R_s=3R_b$. Note that in
Eq.~(\ref{eq:lang5}) all the parameters are known. Clearly, the
agreement is very good. This confirms that even for a large test
particle, the motion is well described by the simple Langevin
equation. This observation, while reasonable, is not trivial since the
limited size of our system and the radius of the particles are
comparable to the mean free path.

We can now present a theoretical calculation of $\gamma$ which
describes dissipation in the Langevin equation for all pairs
($R_s$,$R_b$). This will allow us to compare the theoretical values
with the numerical ones as a function of $R_s$.

The value of $\gamma$ depends on both velocities, $v_s$ and $v_b$.  To
estimate theoretically the value of $\gamma$, we consider the
deviation, due to a collision, of the particle {\bf s} moving at $v_s$
in the $x$ direction. The dissipative term $-\gamma \vec{v_i}$
appearing in Eq.~(\ref{eq:lang1}), in the $x$ direction, can therefore
be formally written as
\begin{equation}
\label{eq:gamma}
-\gamma v_s = <\frac{\vec{v'}_s.\vec{x} -v_s}{v_s}>\omega_c v_s
\end{equation}
where $\vec{v'}_s$ is the velocity after the collision and $\omega_c$
is the rate of collision. The symbol $< >$ in Eq.~(\ref{eq:gamma})
corresponds to the average over all collisions between the {\bf s}
particle and the {\bf b} ones. To calculate the different terms, we
proceed as follows. We consider the collision of {\bf s} with a
particle {\bf b} moving at a velocity $\vec{v}_b$.  The collision is
characterized by two angles: $\theta$, the angle between $(\vec{r_s} -
\vec{r_b})$ and the $x$ axis, and $\varphi$ the angle between
$\vec{v_b}$ and the $x$ axis. Then, for such a collision, illustrated
in Fig.~\ref{fig:colij}, we can calculate theoretically
$\vec{v'_s}(\theta,\varphi)$, the final velocity of the {\bf s}
particle.

For elastic collisions, the projection of $\vec{v'_s}(\theta,
\varphi)$ on the $\vec{x}$ direction is given by,
\begin{equation}
\label{eq:v'sx}
\vec{v'_s}(\theta,\varphi).\vec{x}= \frac{m_s -m_b}{m_s+m_b} v_s
cos^2(\theta) + \frac{2m_b}{m_s+m_b}v_bcos(\theta)cos(\theta -\varphi)
+ v_ssin^2(\theta),
\end{equation}
with the collision taking place only if
\begin{equation}
\label{eq:col}
v_scos(\theta) - v_bcos(\theta -\varphi) > 0.
\end{equation}
Integrating over $\varphi$ and taking into account Eq.~(\ref{eq:col})
we can write
\begin{equation}
\label{eq:v's/fi}
<\vec{v'_s}(\theta).\vec{x}>_{\varphi}=\frac{\int\limits_{0}^{2\pi}{\mathcal
E}_v(v_scos(\theta)
-v_bcos(\theta -\varphi))\vec{v'_s}(\theta,\varphi)\vec{x}d\varphi}
{\int\limits_{0}^{2\pi}{\mathcal E}_v(v_scos(\theta)
-v_bcos(\theta -\varphi))},
\end{equation}
where ${\mathcal E}_v$ is the Heavyside function.  We found for
Eq.~(\ref{eq:v's/fi}) two solutions depending on the velocities.  If
$v_s <v_b$, we have
\begin{equation}
\label{eq:sol1}
<\vec{v'_s}(\theta).\vec{x}>_{\varphi}= \frac{m_s
-m_b}{m_s+m_b}v_scos^2(\theta) + v_s sin^2(\theta) -
\frac{2m_bv_bcos(\theta)sin(\theta_p)}{(\pi - \theta_p)(m_s+m_b)}
\end{equation}
for all $\theta$ ($0 \le \theta \le \pi$) and with $\theta_p = {\rm
cos}^{-1}(v_s{\rm cos}(\theta)/v_b)$.  For the second case, $v_s>v_b$,
there is a critical angle, $\theta_c={\rm cos}^{-1}(v_b/v_s)$, such
that for $\pi - \theta_c < \theta \le \pi + \theta_c$ the collision
does not take place.  In this case the solution of
Eq.~(\ref{eq:v's/fi}) is
\begin{equation}
\label{eq:sol2}
\begin{array}{l}
\displaystyle{<\vec{v'_s}(\theta).\vec{x}>_{\varphi}=\frac{m_s
-m_b}{m_s+m_b}v_scos^2(\theta) + v_s sin^2(\theta) \quad \text{ for}\: 0
\le \theta  < \theta_c},\\
\displaystyle{<\vec{v'_s}(\theta).\vec{x}>_{\varphi}= \frac{m_s
-m_b}{m_s+m_b}v_scos^2(\theta) + v_s sin^2(\theta) -
\frac{2m_bv_bcos(\theta)sin(\theta_p)}{(\pi - \theta_p)(m_s+m_b)}
\quad \text{ for} \: \theta_c \le \theta \le \pi - \theta_c}.
\end{array}
\end{equation}
We call $\nu(\theta)$ the mean relative loss of velocity,
$\nu(\theta)=\frac{<\vec{v'}_s(\theta).\vec{x}>_{\varphi} - v_s}{v_s}$,
averaging only over the angle $\varphi$.  In Fig.~\ref{fig:nu-theta}
we show $\nu(\theta)$ for the particular case $R_s=0.25$ and
$R_b=0.5$, which means that $v_s > v_b$. Note that in our calculation,
the terms $v_s$ and $v_b$ correspond to the averaged values with
respect to the appropriate Maxwellian distribution.  To obtain the
mean value, $\tilde{\nu}$, of $\nu$, we average by integrating
numerically over $\theta$.

To conclude the calculation of the dissipative term, $-\gamma v_s$, we
have to estimate, using Eq.~(\ref{eq:gamma}), the collision frequency
which also depends on the velocities of the two particles. A similar
calculation of $\nu$ can be done~\cite{henr}. In the stationary state
where $v^2_s$ and $v^2_b$ are constant and the distributions of the
velocities are Maxwellian one can use~\cite{chapman}
\begin{equation}
\label{eq:wc}
\omega_c= \chi\sqrt{\pi}(R_s + R_b)d\sqrt{v^2_s + v^2_b},
\end{equation}
where $d$ is the density of {\bf b} particles and $\chi$ is a
correction factor which corresponds to the local radial distribution
around the {\bf s} particle.

In Fig.~\ref{fig:D-el}, we compare, for different values of $R_s$, the
diffusion coefficient found from the simulation with the theoretical
value, $v^2(\infty)/2\gamma$, predicted by the Langevin equation
combined with our analytical calculation of $\gamma$.  The theoretical
calculation of $\gamma$ agrees very well with the simulation
results. We recall that $1/\gamma$ corresponds to the characteristic
time for the diffusive behaviour.  It is important to notice that
$\gamma$ can be approximated by $\omega_c$ only when $R_s <<
R_b$. Effectively, the calculation for $m_s \sim 0$ gives $\tilde{\nu}
= -1$. Larger {\bf s} particles need to suffer more than one collision
to lose memory of their previous condition. For $m_s \sim
\infty$, $\tilde{\nu}$ is found equal to zero.  Using these methods,
we find for the elastic monodisperse case ($R_s=R_b$) that
relaxation (decorrelation) takes place after about three collisions.

The agreement between numerical results and theoretical predictions
allows us to confirm our numerical algorithm.

\section{Forced system}

\label{therm}
In a real granular system dissipation occurs through collisions, a
fact that must be taken into account. Experimental mechanical
properties of grains (restitution and friction coefficients) and
collision laws~\cite{foerster} are used in our simulations. The
collisions between grains and the walls are treated with the same
inelastic properties.  Due
to dissipation, we need to feed energy into the system to maintain the
particles agitated.  To accomplish this, we choose random
heating~\cite{noije,will}: At every time step $\delta t$ we give a
random acceleration, $\eta_i(t)$, in both spatial directions to each
particle. The equation of motion can now be written formally as:
\begin{equation}
\begin{array}{c}
m\frac{d\mathrm{v}_i}{dt}= F_i^c + F_i^t,\\
<F_{i}^t(t)F_{j}^t(t')>= m^2\delta_{i,j}\delta(t-t') \eta_0^2.
\end{array}
\label{eq:dyn2}
\end{equation}
$F_i^c$ is the collision force acting on a particle of mass $m$. We
chose the random acceleration, $F_i^t/m$, to be independent of the
mass of the particle. It is given by a Gaussian noise of variance
$\eta_0^2$.

At long time, the loss of energy due to collisions and the gain due to
$F^t$ balance each other such that the system reaches a steady state
out of equilibrium. It can be shown~\cite{peng} that the velocity
distribution in this steady state is well described by a Maxwellian.

\subsection{Stationary state}

In the stationary state energy loss and gain balance
exactly.  The energy loss per unit time, $\Gamma$, for the {\bf s}
particle, can be expressed as:
\begin{equation}
\Gamma = P(m_s,m_b)  \omega_c m_s v^2,
\end{equation}
where $P(m_s,m_b)$ is the relative energy loss of particle {\bf s} due
to collisions.  Clearly as for $\tilde{\nu}$, $\Gamma$ must depend on
the mass of the particle and on the two velocities $v_s$ and $v_b$.
On the other hand, the gain of energy due the stochastic force is
\begin{equation}
\frac{1}{2}m_s[v^2(t+\delta t) - v^2(t)]=m_s\eta_0^2\delta t.
\end{equation}
In the steady state of the monodisperse system ($R=R_s=R_b$, and
$v^2(\infty)=$constant), we find, using Eq.~(\ref{eq:wc}), the
following scaling for $v^2(\infty):$
\begin{equation}
\begin{array}{c}
v^2(\infty) \propto (\eta_0^2)^{2/3} \\
v^2(\infty) \propto \tau_c 
\end{array}
\label{eq:stat2}
\end{equation}
We checked these two scaling laws numerically (see
Fig.~\ref{fig:v2-more}) and obtained the correct exponent $2/3$ for
the various coefficients of restitution used in the contact laws. We
have also verified the predicted dependence on $\tau_c$ for different
values of $R$. The good agreement between theory and simulation
indicates that we can describe the system by macroscopic continuous
equations if $\delta t << \tau_c$.  As we explain
elsewhere~\cite{mixture}, the term $P(m,m)$ (in the monodisperse case)
is independent of mass and velocity, because all particles are
identical. This value of $P(m,m)$ was found equal approximatively to
0.145 for the mechanical properties corresponding to acetate
spheres~\cite{foerster}.  We can then, in the case of a mono-disperse
system, predict the dependence of the mean square velocity on the
various parameters and, consequently, characterize the stationary
state.  For the bi-disperse case, the calculation is more
complicated. Indeed the loss of energy depends on the two types of
colliding particles and also on the different coefficients of
restitution and friction introduced in the collision laws. As we
show~\cite{mixture} the dependence of $P(m_s,m_b)$ on $v_s/v_b$ is not
trivial.

In this paper we limit ourselves to the effect of the thermalization
mode (or random force) on the diffusion coefficient. To this end, we
will compare in the following section the simulation results for $D$
with $v^2(\infty)/2\gamma$ from the Langevin equation.

\subsection{Diffusion of one particle}

To estimate $D$, we use reflecting boundaries and the same method
explained in section~\ref{sec:reflected}.  We consider here the
bi-disperse case (a single particle of radius $R_s$ in a sea of
particles of radius $R_b$).  As we have not yet found a theoretical
expression for $P(m_s,m_b)$ for this case, we use for the mean square
velocities the values obtained from the simulations which are shown in
Fig.~\ref{fig:res-ine}a.  Note that $\eta_0^2$ has been chosen such
that the value of $v^2_b$ is the same as in the previous section. We
see that $v^2(\infty)$ first decreases with $R_s$ for $R_s < R_b$ but
then increases when $R_s>R_b$. Because of dissipation and the random
acceleration, the repartition of the energy with the mass is no longer
proportional to $1/m_s$.  In all cases it is possible to calculate the
mean collision frequency for {\bf s} with Eq.~(\ref{eq:wc}) and the
associated $\gamma$ value with Eq.~(\ref{eq:gamma}).  We can then
calculate the relaxation time $\tau_r$ for all couples ($R_s$,$R_b$)
used.  In Fig.~\ref{fig:res-ine}b we show the diffusion coefficient
$D$, and the relaxation time $\tau_r$ -in the inset- as functions of
$R_s$.  The behavior of $v^2(\infty)$ strongly modifies the curve of
$\tau_r$ and $D$ versus $R_s$ compared to the elastic case. Note that
the relaxation time represented in Fig.~\ref{fig:res-ine} clearly
increases as $R_s$ increases.
 
We have seen in the elastic case that {\small
$D=\frac{v^2}{2\gamma}$}.  In Fig.~\ref{fig:res-ine}b, we show the
numerical results for $D$ as a function of $R_s$ and the corresponding
values given by {\small $\frac{v^2}{2\gamma}$}. One can see clearly
that the external noise modifies the dynamics of the granular gas and
in particular the diffusion coefficient, $D$. The numerical value of
$D$ is found to be larger than that obtained by the corresponding
random walk.  Indeed, at short time, due to the random force, $v^2(t)$
is not constant. Between two collisions $v^2(t)$ increases linearly
with $t$. Starting with the equation of motion of particle {\bf s}
(between two collisions),
\begin{equation}
\label{eq:dyn} 
\frac{d\mathrm{v_i}(t)}{dt} = \eta_i(t),
\end{equation}
and with the initial conditions $x_i(0)$ and $v_i(0)$, we can
calculate mean square displacement
\begin{equation}
<(x_i(t)-x_i(0))^2> =
\left <\int\limits_{t_1=0}^{t}dt_1 \left( \mathrm{v}_i(0) +\int\limits_{0}^{t_1}
\eta_i(t'_1)dt'_1 \right) \int\limits_{t_2=0}^{t}dt_2 \left(\mathrm{v}_i(0) +
\int\limits_{0}^{t_2} \eta_i(t'_2)dt'_2 \right) \right>.
\end{equation}
In two dimensions, this yields for the interval between two collisions
\begin{equation}
\label{eq:r2-ine}
<(r(t)-r(0))^2)> = v^2(0) t^2 + \frac{2\eta_0^2}{3}t^3.
\end{equation}
On the other hand, in the case of a random walk (or elastic
collisions) the mean square displacement at short time scales as
$t^2$.  This difference explains the disagreement between $D$ and
$\frac{v^2(\infty)}{2\gamma}$.  As the velocity changes between two
collisions the probability of collision is increasing with time
too. The calculation of the coefficient of diffusion is not easy in
this case, due to the correlation between the velocity and the
probability of collision (see Eq.~\ref{eq:wc}).  For very small
particles, if one approaches relaxation by the time of a new
collision, {\it i.e.} $\tilde{\nu} \simeq -1$, this calculation should
be possible. Indeed we can assume that the velocities before and after
a collision are not correlated and have the same distribution (molecular
chaos). We can then compute the mean square displacement, knowing the
dependence of the collision probability on the
velocity~\cite{henr}. With this assumption we improve the estimate of
$D$ for the smallest $R_s$. But for bigger particles we have seen that
the velocities stay correlated over many collisions and we can no
longer use molecular chaos.

\section{Conclusions}

We have presented here some general results about the diffusion
process in an agitated granular gas.  We first showed that the
boundary conditions used in the simulations are of crucial importance.
Indeed, periodic boundary conditions introduce artificially strong
temporal correlations which alter the macroscopic properties of the
gas. If we ensure that no correlations are induced by the algorithm,
for example by using reflecting boundaries, the numerical results
obtained for an elastic gas can be described very well by a Langevin
equation. We have presented a theoretical calculation of the
relaxation time which allows us to predict the diffusion coefficient
in all cases studied. This was not a priori intuitive since the radius
of the particles is of the order of the mean free path. Finally we
have analyzed the influence of uniform heating (a random acceleration)
on dissipative gases. We have shown that heating influences the
dynamics at short time. This is evident through the value of the
diffusion coefficient which is different from that expected from the
Langevin description. We are now applying with success these results
to the diffusion process in a granular mixture consisting of two type
of grains (differing by mass or size) in equal proportion.

\vspace{2ex}

\centerline{\bf ACKNOWLEDGMENTS}
\vskip 0.5cm

This work was partially funded by the CNRS Programme International de
Cooperation Scientifique PICS $\#753$ and the Norwegian research
council, NFR.

%%%%%%%%%%%% figure caption %%%%%%%%%%%%%%%%%%%

\begin{figure}
%\centering
%\includegraphics[height=4.5cm]{perio2.eps} \hspace{0.5cm}
%\includegraphics[height=4.5cm]{corel-ped.eps}
\caption{Dependence of the mean square displacement on the system
size.  {\bf (a)}: $<(\vec{r}(t+t_0) - \vec{r}(t_0)>$ as a function of
$t$. From bottom to top the system size is 20, 40, 30, 50, 60.  {\bf
(b)}: Integral of $C(t)$ as a function of $t$ for the same
system. From bottom to top the system size is 20, 40, 30, 50, 60,
respectively}
\label{fig:perio}
\end{figure}

\begin{figure}
\caption{$D_1$ and $D_2$ vs time for a typical monodisperse case with
periodic boundaries.}
\label{fig:D1D2}
\end{figure}

\begin{figure}
\caption{illustration of the non-conservation of the distance between
two particles.} 
\label{fig:iandj}
\end{figure}

\begin{figure}
\caption{ Total angular momentum $L_z(t)$ vs $t$ for two system lengths
and the same particle density. (Full line): $L= 20$, (Dashed line):
$L= 50$.}
\label{fig:lz-t}
\end{figure}  

\begin{figure}
%\centering
%\includegraphics[width=5cm]{R2-wall.eps}
\caption{$<(\vec{r}(t)-\vec{r}(0))^2>$ as function of $t$
using reflecting boundaries. Superposed (as in figure
\ref{fig:perio}) are the results for system sizes: 20, 30, 40, 50, 60.}
\label{fig:R2_wall}
\end{figure}

\begin{figure}
\caption{A particle of mass $m_s$ colliding with a particle of mass
$m_b$: definition of the angles $\theta$ and $\varphi$.}
\label{fig:colij}
\end{figure}

\begin{figure}
%\centering
%\includegraphics[width=4.5cm]{v-t_lange.eps} \hspace{0.5cm}
%\includegraphics[width=4.5cm]{r2-t_lange.eps}
\caption{Comparison between numerical results and Langevin
approximation.  $R_s=1.5$ and $R_b=0.5$. {\bf (a)} dependence of the
mean squared velocity on $t$; $\bigcirc$: numerical result,
full line: Fit using $e^{-2\gamma t}$ according to
Eq.~(\ref{eq:lang3}). {\bf (b)} the mean squared displacement,
$\bigcirc$: numerical result, full line: Theoretical prediction
according to Eq.~(\ref{eq:lang4}) using for $\gamma$ the value
obtained from Fig.~\ref{fig:lange}a.}
\label{fig:lange}
\end{figure}

\begin{figure}
\caption{ $\nu(\theta)$, the mean relative loss of energy per collision in
the $\theta$ direction, for $R_s=0.25$, $R_b=0.5$ and $v^2_b=25$}
\label{fig:nu-theta}
\end{figure}

\begin{figure}
%\centering
%\includegraphics[width=5cm]{D-Rs_elast.eps}
\caption{Coefficient of diffusion for different values of
$R_s$. $R_b=0.5$.  $\bigcirc$: Numerical values obtained by
simulation. Full line: theoretical values calculated from
$\frac{v^2(\infty)}{2\gamma}$}.
\label{fig:D-el}
\end{figure}

\begin{figure}
%\centering
%\includegraphics[width=4.7cm]{v2-eta0.eps} \hspace{0.5cm}
%\includegraphics[width=4.5cm]{v2-tc.eps}
\caption{{\bf (a)} $v^2(\infty>$ versus $\eta_0^2$ for different
coefficients of restitution. $\bigcirc$: $e_n =0.87$, $e_s=0.4$,
$\mu=0.25$; $\triangle$: $e_n =0.4$, $e_s=0.4$, $\mu=0.25$. {\bf (b)}
$v^2(\infty)$ versus the mean time between collisions $\tau_c$.}
\label{fig:v2-more}
\end{figure}

\begin{figure}
%\centering 
%\includegraphics[width=4.5cm]{V2-Rs_inelast.eps} \hspace{0.5cm}
%\includegraphics[width=4.7cm]{D_inelast.eps}
\caption{{\bf (a)} $v^2(\infty)$ vs $R_s$ ($R_b=0.5$).  {\bf (b)}
Coefficient of diffusion $D$ as a function of $R_s$. $\bigcirc$:
Numerical values obtained from simulation. Full line: Corresponding
values given by $\frac{v^2(\infty)}{2 \gamma}$.  The insert shows
$\tau_r$ versus $R_s$}
\label{fig:res-ine}
\end{figure}
\end{document}